\documentclass[11pt,onecolumn,superscriptaddress,nofootinbib,floatfix]{revtex4-2}

\pdfoutput=1
\usepackage{amsmath}
\usepackage{amsfonts}
\usepackage{amssymb}
\usepackage{mathrsfs}
\usepackage{graphicx}
\usepackage{color}
\usepackage{bm}
\usepackage{blindtext}
\usepackage{wasysym}
\usepackage{hyperref}
\hypersetup{colorlinks=true,allcolors=blue}
\usepackage[normalem]{ulem}
\usepackage{lineno}

\usepackage{subcaption}

\newcommand{\mnras}{Mon. Not. Roy. Astron. Soc.}

\newcommand{\bi}{\begin{itemize}}
	\newcommand{\ei}{\end{itemize}}

\begin{document}
	
	\title{KM3NeT upper bounds of detection rates of solar neutrinos from annihilations of dark matter at the solar core}
	\author{Aman Gupta}
	\email{aman.gupta@saha.ac.in}
	\author{Debasish Majumdar}
	\email{debasish.majumdar@saha.ac.in}

	\affiliation{Theory Division, Saha Institute of
		Nuclear Physics,
		1/AF
		Bidhannagar,
		Kolkata 700064, India}
	\affiliation{Homi Bhabha National Institute, Anushakti Nagar, Mumbai 400094, India}
	\author{Ashadul Halder}
	\email{ashadul.halder@gmail.com}
	\affiliation{Department of Physics, St. Xavier’s College,30, Mother Teresa Sarani, Kolkata-700016, India.}
	
	\begin{abstract}
		The Weakly Interacting Massive Particles (WIMPs) so far remain one of the most popular candidates for dark matter. If captured gravitationally inside the core of the Sun, these WIMPs may produce high energy neutrinos as the end product in case they undergo self annihilations at the solar core. In this work, we address the detectability of such neutrinos at the proposed KM3NeT detector. Upper bounds of the detection rate for such neutrinos at KM3NeT are computed for the case of a generic dark matter scenario and also when specific models for particle dark matter are chosen. In this work, upper bounds of muon event rates for different annihilating dark matter masses are computed for each of the cases of dark matter annihilation channels (e.g.  $b\bar{b}~, W^+W^-, Z\bar{Z} $ etc). These upper bounds are also computed by considering the dark matter scattering cross-section using upper bounds obtained from PandaX-4T direct dark matter search experiment. 
		
	\end{abstract}
	\maketitle
	
	\section{Introduction}
	The nature of dark matter is still an enigma. The direct detection experiments so far, have not yielded any convincing result for dark matter detection. At the same time there are efforts for indirect detection of dark matter whereby the possible Standard Model end products of dark matter annihilation are probed. The annihilation of dark matter inside a massive astrophysical object may produce Standard Model particles such as $e^+$, $e^-$, $\mu^+$, $\mu^-$, $\gamma-$ photons, $\nu$, $\Bar{\nu}$ etc. In case, dark matter annihilation inside a massive compact astrophysical object produces neutrinos then those neutrinos may possibly be detected by various earth bound neutrino telescopes \cite{Bose_2022,Zentner:2009is,Aartsen2017,Fargion_1998}. Different neutrino telescopes are designed for detection of neutrinos of different energy ranges. Therefore, detection of such neutrinos in one or several telescopes would throw light on the nature and properties of dark matter. \\
	One such upcoming telescope namely KM3NeT \cite{arxiv.2203.10048,Aiello_2021,Aiello_2022,Assal_2021,Sinopoulou_2021,Aiello_2019} is installed in the Mediterranean  sea and uses the sea water as a detecting material. The energy range for KM3NeT is 1-100 TeV and it is sensitive for the energies that may be relevant for dark matter detection.\\
	In this work, we address the detectability of neutrinos in case they originate from dark matter annihilation in the solar core. We have considered first a generic WIMP dark matter and explored the upper bound of the neutrino detection rates by KM3NeT for each of the possible annihilation channels that the dark matter may undergo to produce neutrinos as the final product. We also consider a particle dark matter in the framework of a particle physics model model namely Inert Doublet Model (IDM) \cite{doi:10.1142/S0217732308025954, LopezHonorez:2010eeh} and obtain such bounds with this dark matter candidate. Then we extend our analysis for the case of a super-symmetric (SUSY) dark matter particle namely neutralino within the framework of Minimal Supersymmetric Standard Model (MSSM) \cite{doi:10.1142/9789812839657_0001, Gunion_2006} and elaborately study the detectability of such neutralino dark matter via indirect detection in case these neutralinos annihilate at the solar core to produce neutrinos as end product. \\
	The paper is organised as follows. In section 2, we provide a brief account of formalism used in this work. We also describe the possible annihilation channels through which dark matter may annihilate. In section 3, we furnish the calculations and the results of our analysis. Finally in section 4, we summarize the work with some discussions.  
	
\section{formalism} 
    The differential flux of $i^{\rm th}$ flavour of neutrinos ($i = \nu_\mu, \bar{\nu}_\mu$), as would be detected by earth-based detectors and which are produced in the Sun due to the WIMP annihilation process are given by \cite{jungman}
\begin{equation}
	\left( \dfrac{d\phi}{dE} \right)_i =\dfrac{\Gamma_A}{4\pi R^2} \displaystyle\sum_F B_F \left( \dfrac{dN}{dE} \right )_{F,i},
\end{equation}
where $R$ is distance between the Earth and the Sun, $B_F$ represents the branching for the annihilation channel $F$ and $\left( \frac{dN}{dE} \right )_{F,i}$ is the differential neutrino spectrum for the $i^{\rm th}$ flavour and the annihilation channel $F$. In the above expression, $\Gamma_A$ denotes the total rate for WIMP annihilation in the Sun, which essentially depends on total number of dark matter particle captured gravitationally in the Sun ($N$) as \cite{Zentner:2009is},
\begin{equation}
	\Gamma_A =C_{\rm ann} N^2,
	\label{eq:Gamma}
\end{equation}
where $C_{\rm ann}$ is the rate per pair at which the captured DM particles annihilate within the Sun. The total number of dark matter particle captured ($N$) can be expressed as,
\begin{equation}
	\dfrac{dN}{dt} = C_C + (C_{SC} - C_{SE}- C_E)N -\left(C_{\rm{ann}} + C_{\rm{sevap}}\right) N^2,
	\label{eq:dNdtgen}
\end{equation}
where $C_C$ is the rate of dark matter capture due to nuclei-DM elastic scattering, while $C_E$ denotes the same at which the captured dark matters evaporate by scattering. Similarly, $C_{SC}$, $C_{SE}$ and $C_{\rm{sevap}}$ are the rate for DM self-capture, self-ejection and self-evaporation, respectively \cite{Gaidau:2018yws,Chen:2014oaa}. 
We assume a Tsallis velocity distribution function $f^\prime (v)$ of speeds $v$, which is the generalized form of the Maxwell-Boltzmann distribution. It is expressed as \cite{Nunez-Castineyra:2019odi}
\begin{align}
     f^\prime(v) = \dfrac{1}{N(v_0,q)}\Bigg(1-(1-q)\dfrac{v^2}{v_o^2}\Bigg)^{q/(1-q)}
	\label{eq:halosunframe}
\end{align}
where $N(v_0,q)$ is the normalization factor. The value of $v_0$ and $q$ is adopted from Fig. (5)a of Ref. \cite{Nunez-Castineyra:2019odi}  and given by $245.9$ km/sec and $0.864$, respectively. These values are obtained by fitting the simulated velocity distribution with the Eq. \ref{eq:halosunframe}. We then obtain the velocity distribution function in the frame of reference of Sun $f(u)$, by using the following conversion formula.  
\begin{align}
    f(u) = \int_{-1}^{1} f\Bigg(\sqrt{(v^2+\tilde{v}^2 +2v\tilde{v}\cos{\theta}} )\Bigg)d\cos{\theta}
    \label{eq:f_u}
\end{align}
where $\Vec{u} = \Vec{v}+\Vec{\tilde{v}}$ and $\theta$ is the angle between them. According to recent estimations, the average speed of dark matter ${v}=288$ km/s \cite{2009ApJ...704.1704B,2009ApJ...700..137R,2010MNRAS.402..934M} and average velocity of the Sun with respect to galactic dark matter halo $\tilde{v}=247$ km/s \cite{2010MNRAS.403.1829S}.

So, the nuclear capture rate per unit volume at a distance $r$ from the Solar core due to species $i$ can be written as \cite{1987ApJ...321..571G}
\begin{equation}
	\frac{dC_{\text{c},i}}{dV} = n_{\chi} \int_0^{\infty}  \frac{w}{u}  \Omega_{i}(w) f(u) du,
	\label{eq:dccidv}
\end{equation}
where $n_{\chi}=\rho_{\odot}/m_{\chi}$ is the number density of DM particle in galactic halo near Sun.
In the above expression, the instantaneous speed of dark matter particles $\omega$ depends of the escape velocity $v_{\rm {esc}}$ as $\omega^2=u^2+v_{\rm {esc}}^2$ and capture rate of DM particles at velocity $\omega$ is given by \cite{Zentner:2009is},
\begin{equation}
	\Omega(\omega) = n_i \sigma_i v_{\rm {esc}}\dfrac{v_{\rm {esc}}}{\omega}\left[1-\dfrac{u^2}{v_{\rm {esc}}^2}\dfrac{(m_{\chi}-m_i)^2}{4m_{\chi}m_i}\right],
	\label{eq:Omegaomega}
\end{equation}
where $n_i$ is the number density of the species $i$ and $m_i$ is the mass of the scattered particle. The cross-section $\sigma_i$ in Eq. \ref{eq:Omegaomega} denotes the DM-nucleus scattering cross-section that finally results in the capture of dark matter in the Sun \cite{Zentner:2009is}. Needless to mention that DM-nucleus scattering cross-section can be derived from DM-nucleon scattering  cross-section $\sigma_P$.

The dark matter capture coefficient by nuclei, $C_C$ can be obtained by integrating over the entire volume of the Sun and further summing over the effects of all constituents, (i.e. Hydrogen, Helium, Carbon, Nitrogen, Oxygen, Iron etc.) given by
\begin{equation}
	C_C = \sum_{i}\left(\int_0^{R_{\odot}}\frac{dC_{C,i}}{dV}d^3r\right) = \sum_{i}\left(\int_0^{R_{\odot}}4\pi r^2 \frac{dC_{C,i}}{dV}dr\right).
	\label{eq:cc}
\end{equation}

\begin{figure}[htp]
\centering
\begin{subfigure}[b]{0.5\linewidth} 
\centering
\includegraphics[width=1.0\linewidth]{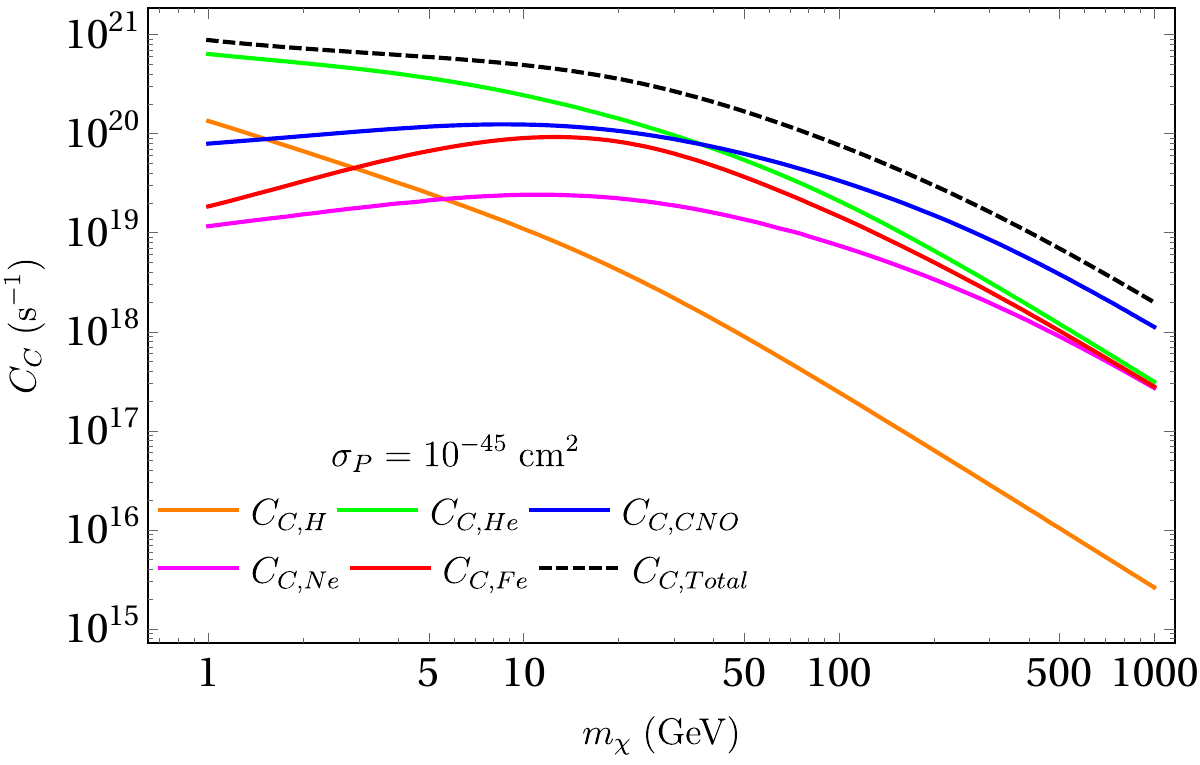} 
\subcaption{} 
\end{subfigure}\hfill
\begin{subfigure}[b]{0.5\linewidth} 
\centering
\includegraphics[width=1.0\linewidth]{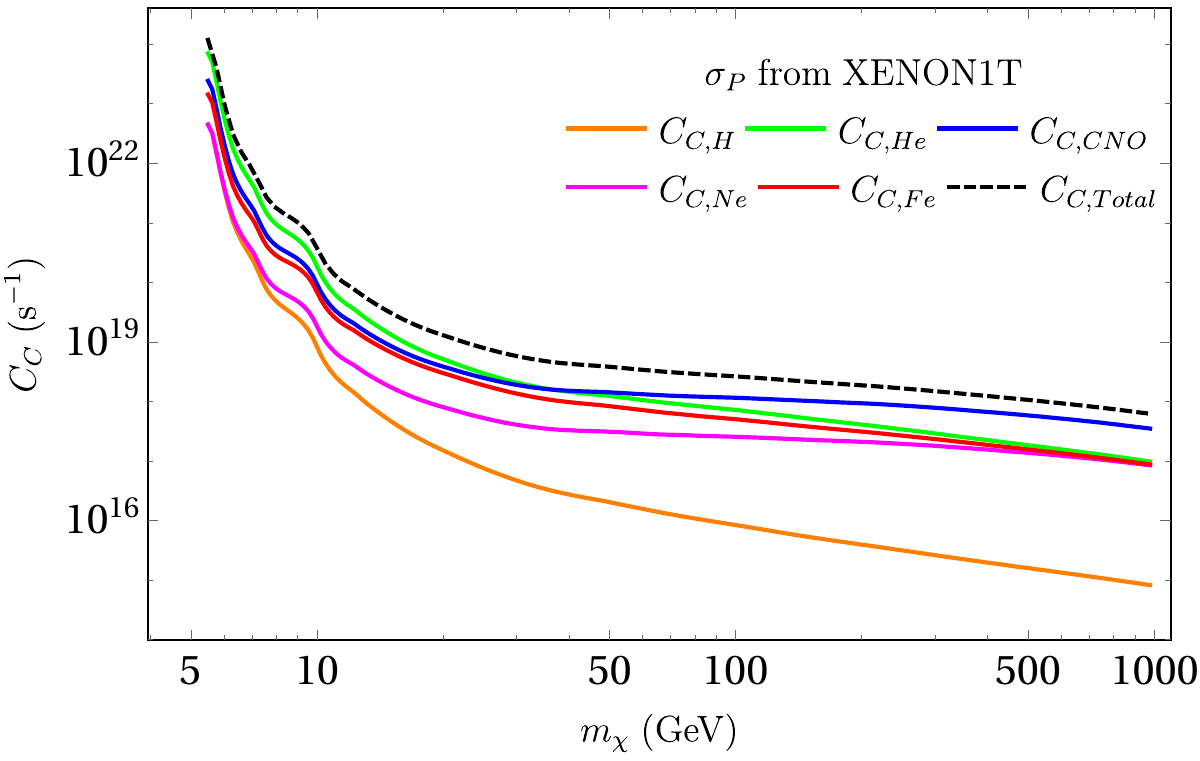} 
\subcaption{} 
\end{subfigure}
\vspace{9pt}
\begin{subfigure}[b]{0.5\linewidth} 
\centering
\includegraphics[width =1.0\linewidth]{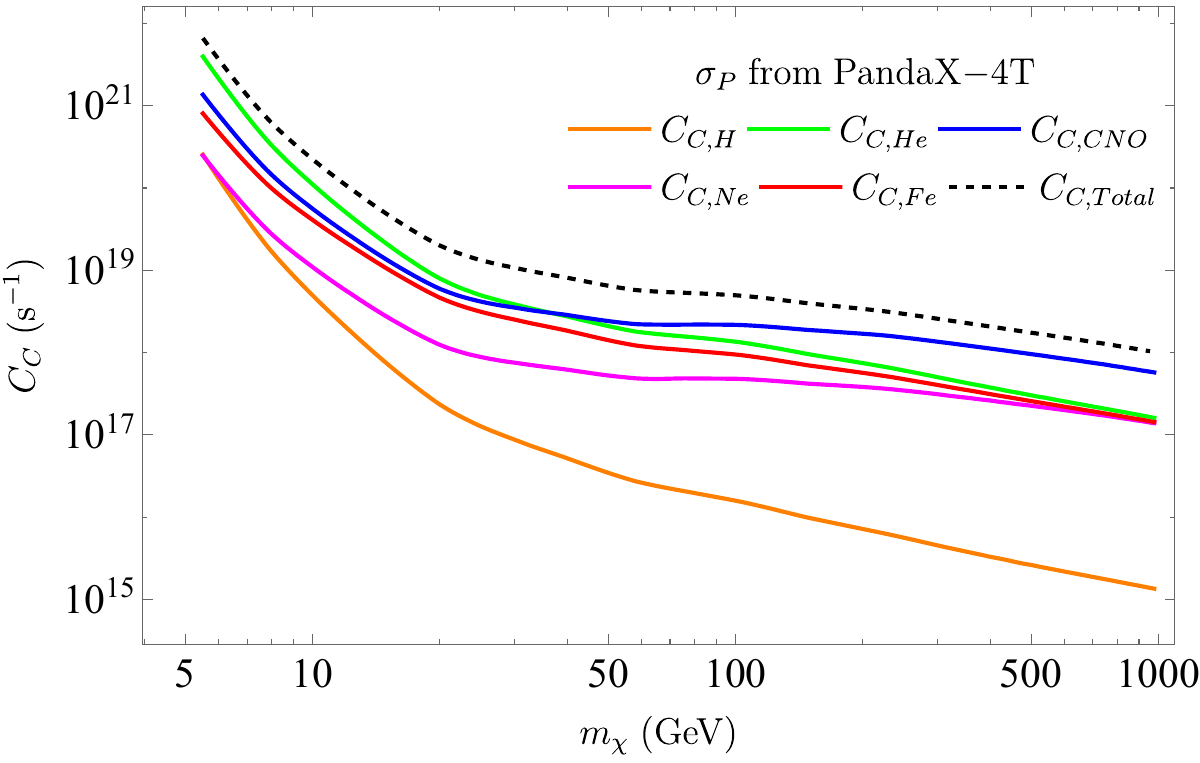} 
\subcaption{} 
\end{subfigure} 
\caption{Variation of $C_C$, the dark matter capture co-efficient by nuclei, in the Sun with DM mass $m_{\chi}$. In Left panel constant value of $\sigma _P ~(10^{-45} \text{cm}^2)$ is adopted. On the other hand, in right panel $ \sigma_p$ is taken from the upper bound of the cross-sections given by XENON1T experiment while in the lower-middle panel we have used latest results of PandaX-4T direct detection experiment.  In all plots, the orange and green lines denote the capture due to hydrogen and helium,  respectively. The blue lines correspond to combined rate by carbon, nitrogen and oxygen. The red and magenta lines represent the same for iron and neon, respectively. The black dashed line denotes the total capture rate ($C_C$). See text for more details.} 
\label{fig:Cc}
\end{figure}

In Fig.~\ref{fig:Cc}(a), we furnish the dark  matter captured rate inside the Sun driven  by each of the major Solar processes, such as $p-p$ chain, CNO cycle etc. The total capture rate is also shown in Fig. \ref{fig:Cc}(a). In computing $C_c$ using the above equations, a constant value for $\sigma_p$ (=$10^{-45}$ cm$^2$) has been chosen. This is to be mentioned here that we have adopted the Solar density profile of individual elements from a new generation of standard solar models \cite{Vinyoles:2016djt}. The computations of $C_c$ are also done with the $\sigma_p$  value adopted from the upper bounds of XENON1T \cite{Aprile_2018} and PandaX-4T \cite{PandaX-4T:2021bab} direct detection experiments. These results are plotted in Fig. \ref{fig:Cc}(b) and Fig. \ref{fig:Cc}(c). It is to be noted that in obtaining the results shown in Fig. \ref{fig:Cc}(b)  we use the upper limit of dark matter-nucleon scattering cross-section corresponding to each dark mater mass as given by XENON1T and PandaX-4T experiment. On comparing the Fig. \ref{fig:Cc}(a) and Fig. \ref{fig:Cc}(b) we observe that the DM capture rate for XENON1T is about $10$ times higher than that of PandaX-4T.

\begin{figure}
	\centering
	\includegraphics[width=0.7\linewidth]{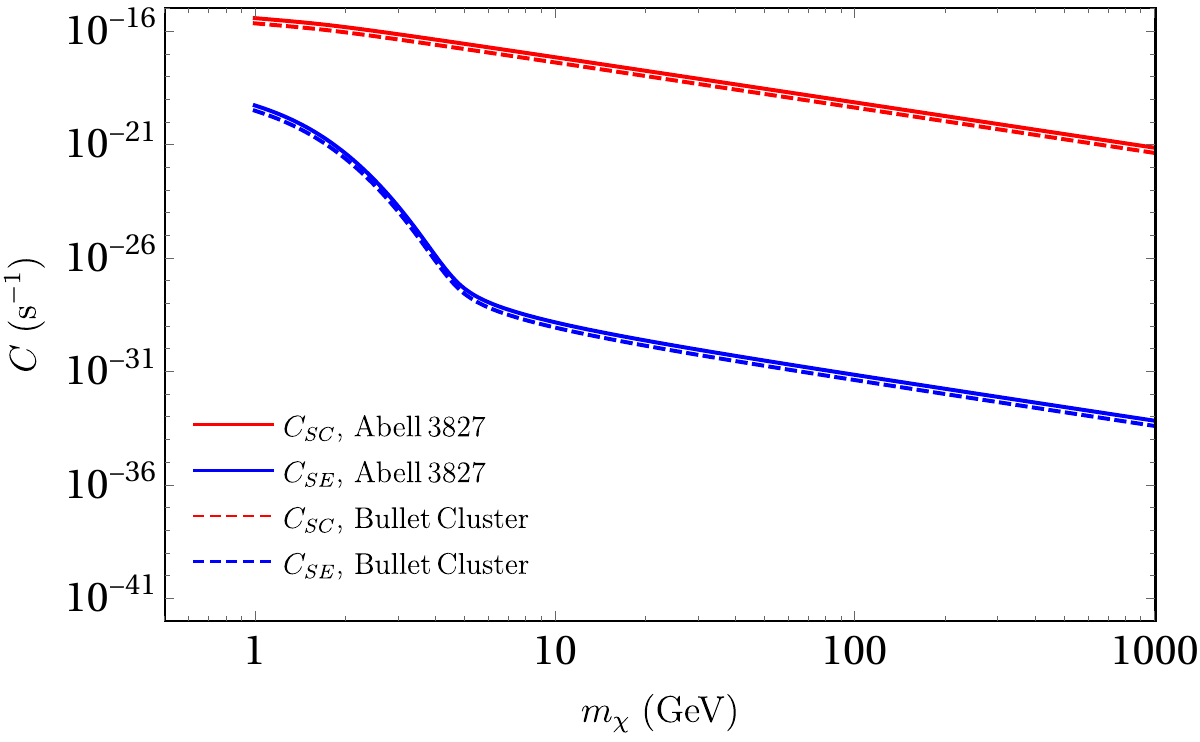}
	\caption{Self-capture (Red) and self-ejection (Blue) rates in the Sun as a function of dark matter mass $m_{\chi}$, for self-interaction cross-sections $\sigma_{\chi\chi}$ obtained from observational limit of bullet cluster (dashed line) and Abell 3827 (solid line).}
	\label{fig:csc_cse1}
\end{figure}

The estimations of the capture and ejection rate of DM particles due to self-interaction namely $C_{SC}$ and $C_{SE}$, respectively, can be estimated in the same way as that for $C_C$. In this case, the cross-section, mass and number density terms will be replaced by the DM self-interaction cross-section ($\sigma_i\rightarrow \sigma_{\chi\chi}$), dark matter mass ($m_i \rightarrow m_{\chi}$) and the captured dark matter density ($n_i\rightarrow n_c$) \cite{Zentner:2009is}. In the present analysis we adopt the observation limits of DM self-scattering cross-section $\sigma_{\chi\chi}$, which are obtained from the analysis of Bullet Cluster \cite{Randall:2008ppe} (i.e. $\sigma_{\chi\chi}/m_{\chi}<1$ cm$^2$/g) and Abell 3827 (i.e. $\sigma_{\chi\chi}/m_{\chi}<1.7$ cm$^2$/g). The results for the variations of $C_{SC}$ and $C_{SE}$ with dark matter mass $m_{\chi}$  are shown in Fig. \ref{fig:csc_cse1}.

The annihilation rate of the gravitationally captured DM within Sun $C_{\rm ann}$ is directly proportional to the velocity-averaged cross-section of dark matter annihilation $\langle \sigma v \rangle$, given by \cite{Zentner:2009is,Gaidau:2018yws}
\begin{equation}
	C_{\rm{ann}} =\dfrac{1}{2}\langle  \sigma v \rangle \int_0^{R_{\odot}}
	4\pi r^2 A^2\exp\left(\frac{-m_{\chi} \Phi(r)}{T_{\rm{core}}}\right)dr,
\end{equation}
where $T_{\rm{core}}$ is the core temperature of Sun and $\Phi(r)$ is the gravitational potential at a radial distance $r$ from the centre of the Sun. The evaluation of $\langle \sigma v \rangle$ is generally dark matter model dependent.  However, in this work we compute $\langle  \sigma v \rangle$  in a model independent way and it is described in \hyperref[app_sigmav]{Appendix A}. 

\begin{figure}
	\centering
	\begin{tabular}{cc}
	\hspace{-0.7cm}
		\includegraphics[width=0.48\linewidth, height=0.43\linewidth]{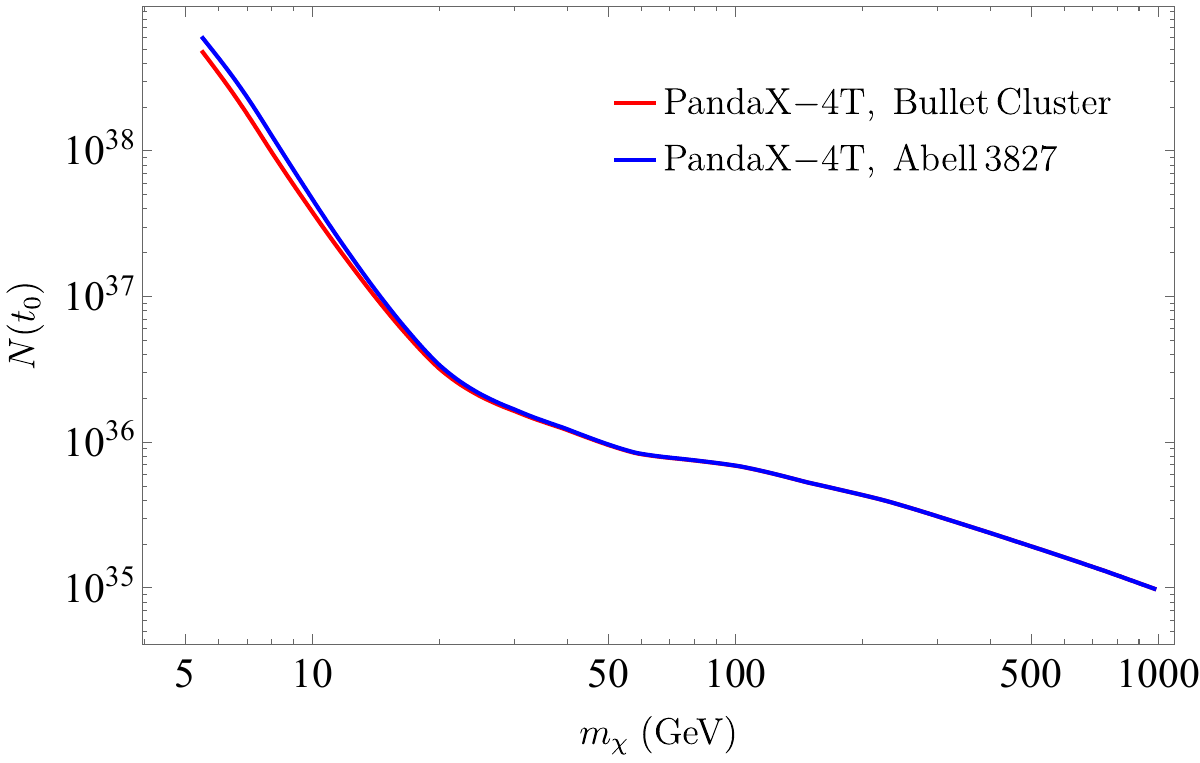}&~
		\includegraphics[width=0.5\linewidth]{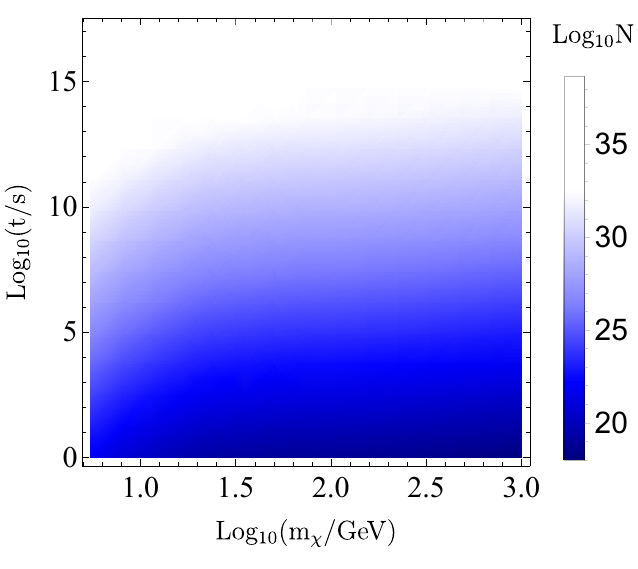}\\
		(a)&(b)\\
	\end{tabular}
	\caption{\label{fig:N_vs_t_mchi} (a) The amount of dark matter containing the Sun at the present epoch ($N(T_0)$) for different dark matter mass $m_{\chi}$. The red line corresponds the case, where the $\sigma_{\chi\chi}$ is obtained from the observational limit of Bullet cluster, while the blue line represents the same for limit of Abell 3827. (b) Contour representation of the total number of dark matter particle captured within the Sun at different time $t$ and dark matter particle mass $m_{\chi}$. In this case $\sigma_{\chi\chi}$ is obtained from the observational limit of Bullet cluster.}
\end{figure}

\begin{figure}
	\centering
	\includegraphics[width=0.7\linewidth]{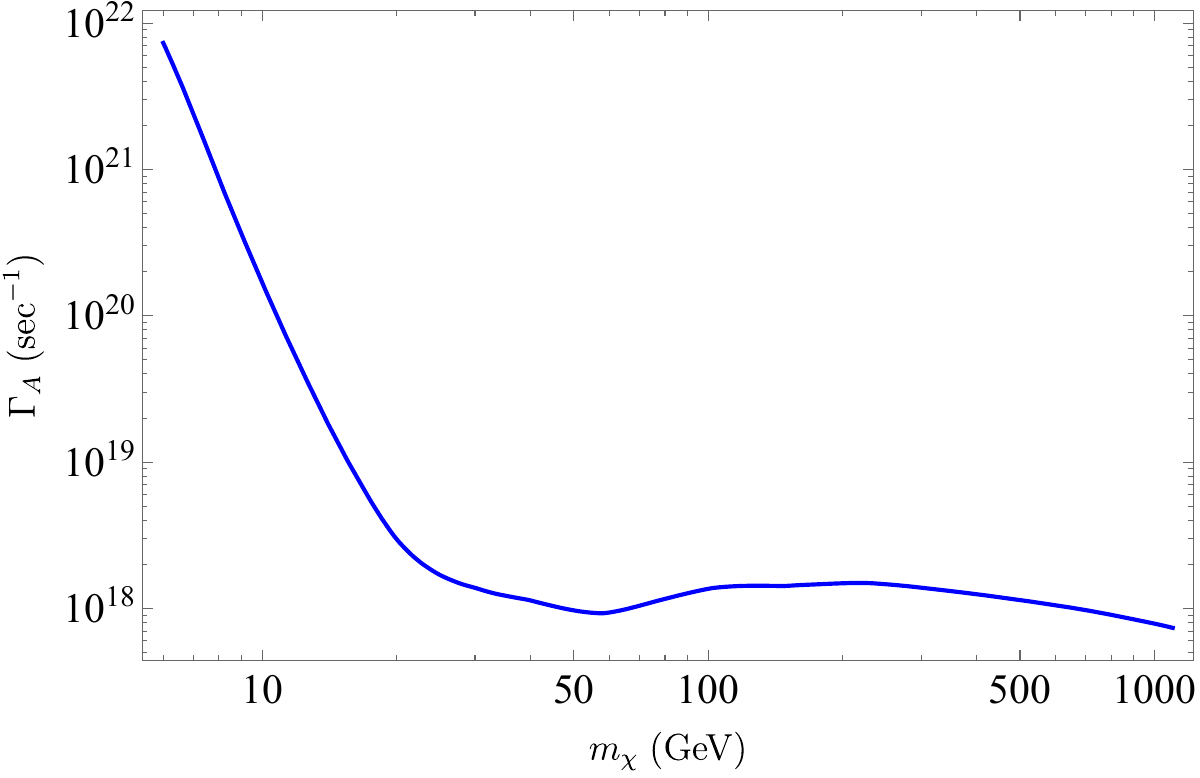}
	\caption{Variation of the total rate for WIMP annihilation in the Sun with DM mass $m_{\chi}$.}
	\label{fig:gamma_a}
\end{figure}

Now considering the initial condition  as $N(0)=0$ at $t=0$, the general solution of  Eq.~\ref{eq:dNdtgen} at a time $t$ can be written as \cite{Zentner:2009is}
\begin{equation}
	N(t)=\dfrac{C_C \tanh \left( \dfrac{t}{\xi} \right)} {\dfrac{1}{\xi}-\dfrac{C_{SC}-C_{SE}-C_E}{2}\tanh\left(\dfrac{t}{\xi}\right)},
	\label{eq:Nt}
\end{equation}
where $\xi^{-1} =\sqrt{{C_C}(C_{\rm{ann}}+C_{\rm{sevap}}) + (C_{SC}-C_{SE}-C_E)^2/4}$. It is to be mentioned that, for higher dark matter masses ($m_{\chi} \gtrapprox 10$ GeV), $C_{\rm{sevap}}\ll C_{\rm{ann}}$ \cite{Gaidau:2018yws}. As a consequence, we neglect the contribution of $C_{\rm{sevap}}$ in our calculation. Moreover assuming near-thermal distribution, the effect of DM evaporation due to scattering ($C_E$) can be ignored for $m_{\chi}\geq4$ GeV \cite{1987ApJ...321..560G,Busoni:2013kaa}. The variation of $N (t)$ with dark matter mass ($m_\chi$) is therefore computed and shown in left panel of Fig. \ref{fig:N_vs_t_mchi} (Fig. \ref{fig:N_vs_t_mchi}(a)). In the right panel of Fig. \ref{fig:N_vs_t_mchi} (Fig. \ref{fig:N_vs_t_mchi}(b)) we estimate the total number of dark matter captured for a given dark matter mass inside the Sun at a particular instant $t$. The total number of captured dark matter particles are shown in  $m_\chi$ vs $t$ plane by colour shades. The WIMP annihilation rate $\Gamma_A$ as 
 a function of dark matter mass $m_\chi$ is shown in Fig. \ref{fig:gamma_a}. It can be observed from Fig. \ref{fig:gamma_a} that the annihilation rate decreases with the increase of DM mass but there is no significant variation of $\Gamma_A$ for DM mass beyond $\sim$ 50 GeV .  

The total detection rate of muon flux at ground based experiments like KM3NeT due to WIMP annihilation in the Sun is given by (following Ref. \cite{jungman})
\begin{equation}
	\Gamma_{\rm detect}= \left(2.54\times10^{-29}  {\rm yr}^{-1} \right) \frac{\Gamma_A}{{\rm sec}^{-1}}\left(\frac{m_\chi}{\rm GeV}\right)^2\displaystyle\sum_{i=\nu,\bar{\nu}} a_i b_i \sum_F B_F \langle {N_Z}^2 \rangle_{F,i} \times A_{\rm eff},
	\label{eqgammadetect}
\end{equation}
where, $A_{\rm eff}$ is the muon effective area (781.59 m$^2$) at the KM3NeT detector. This is obtained by taking the average of $A_{\rm eff}$  for KM3NeT detector. The variation of $A_{\rm eff}$ with energy is obtained from Fig. 19 of Ref.  \cite{Adri_n_Mart_nez_2016}. In the above equation, $a_i$ denotes the neutrino scattering coefficients ($a_\nu = 6.8$, $a_{\bar{\nu}} = 3.1$ \cite{jungman}), while $b_i$ is the amount of neutrino induced muons in the rock ($b_\nu = 0.51$, $b_{\bar{\nu}} = 0.67$ \cite{jungman}). The coefficient ${\langle {N_Z}^2 \rangle}_{F,i}$ is the second moment of the $i^{\rm th}$ flavour neutrino spectrum, in the case of $F$ channel of WIMP annihilation in the Sun. In the present case three channels, namely $\chi\bar{\chi}\rightarrow\tau\bar{\tau}$, $\chi\bar{\chi}\rightarrow b\bar{b}$, and $\chi\bar{\chi}\rightarrow Z\bar{Z}$, are chosen. The quantity ${\langle {N_Z}^2 \rangle}_{F,i}$ for these annihilation channels are as follows \cite{jungman,ritz}.\\

For $\chi \bar{\chi}\rightarrow \tau\bar{\tau}$ channel,
		\begin{equation}
			\left.\langle {N_Z}^2 \rangle_i (E_{\rm inj})\right|_{\tau} \simeq \Gamma_{\tau\to \mu\nu\bar{\nu}} h_{\tau,i}(E_{\rm inj}\tau_i),
		\end{equation}
		where $(i = \nu_\mu,\bar{\nu}_\mu)$, $E_{\rm inj}$ is the injection energy of the decaying WIMP annihilation product inside the Sun, $\tau_\nu (\tau_{\bar{\nu}}) = 1.01\times 10^{-3} (3.8\times 10^{-4})$ GeV$^{-1}$ and the branching ratio $\Gamma_{\tau\to \mu\nu\bar{\nu}} \simeq 0.18$. The stopping coefficients used in the above equation for neutrinos $\nu$ and anti-neutrinos $\bar{\nu}$ are given by \cite{dmsir},
		\begin{eqnarray}
			h_{\tau,\nu_\mu}(y)	&=& \dfrac{4+y}{30(1+y)^4} \nonumber\\
			h_{\tau,\bar{\nu}_\mu}(y) &=& \dfrac{168 + 354y + 348y^2 + 190y^3 + 56y^4 + 7y^5}{1260(1+y)}.
		\end{eqnarray}
\\
For the self-annihilation channel $\chi \bar{\chi}\rightarrow b\bar{b}$, $\langle{N_Z}^2 \rangle$ is expressed as

		\begin{equation}
			\left.\langle {N_Z}^2 \rangle_i (E_{inj})\right|_{b} \simeq \Gamma_{b\to \mu\nu X} \frac{\langle E_d \rangle^2}{E_i^2} h_{b,i}\left(\sqrt{\langle E_d^2 \rangle} \tau_i \right),
		\end{equation}
		where, the branching ratio $\Gamma_{b\to \mu\nu X} = 0.103$. In the above equation $\langle E_d \rangle$ denotes the mean of the hadron energy, which describes the hadronization and the subsequent decay processes of the quarks produced due to WIMP annihilation in the Sun. This is given by,
		\begin{equation}
			\langle E_d \rangle =
			E_c \exp\left(\frac{E_c}{E_0}\right)E_1\left(\frac{E_c}{E_0}\right),
		\end{equation}
		where, $E_c = 470$ GeV \cite{ritz}, $E_{\rm inj}$ is the energy of the injected quarks and $E_0$ is the corresponding initial hadron energy given by $E_0 = Z_f E_{\rm inj}$. $Z_f(=0.73)$ is the quenching fraction for $b$-quarks to account for the energy loss during hadronization given by,
		\begin{equation}
			E_1(x) = \int_x^\infty \frac{e^{-y}}{y}dy.
		\end{equation}
		The term $\langle E_d^2\rangle = E_c (E_0 - \langle E_d \rangle)$. It is to be noted that in the case of $\chi \bar{\chi}\rightarrow b\bar{b}$ channel, the expressions of $h_{b,i}$ remain same as that for the $\chi \bar{\chi}\rightarrow \tau\bar{\tau}$ channel ($h_{\tau,i}$). \\

Finally, for the channels, $\chi \bar{\chi}\rightarrow W^+W^-$ and $\chi \bar{\chi}\rightarrow Z\bar{Z}$,  

		\begin{eqnarray}
			\left.{\langle {N_Z}^2 \rangle}_i (E_{\rm inj})\right|_{W}
			&\simeq &
			\left. \frac{\Gamma_{W\to \mu\nu}}{\beta} 
			\frac{2+2E\tau_i(1+\alpha_i)+E^2\tau_i^2\alpha_i(1+\alpha_i)}
			{E_{\rm inj}^3 \tau_i^3 \alpha_i (\alpha_i^2 - 1)(1+E\tau_i)^{\alpha_i+1}}
			\right|^{E=E_{\rm inj}(1-\beta)/2}_{E=E_{\rm inj}(1+\beta)/2} \\
			\left .{\langle {N_Z}^2 \rangle}_i (E_{\rm inj})\right|_{Z}
			&\simeq &
			\left. \frac{2\Gamma_{Z\to \nu_\mu\bar{\nu}_\mu}}{\beta} 
			\frac{2+2E\tau_i(1+\alpha_i)+E^2\tau_i^2\alpha_i(1+\alpha_i)}
			{E_{\rm inj}^3 \tau_i^3 \alpha_i (\alpha_i^2 - 1)(1+E\tau_i)^{\alpha_i+1}}
			\right|^{E=E_{\rm inj}(1-\beta)/2}_{E=E_{\rm inj}(1+\beta)/2} 
			\label{eqnzz}
		\end{eqnarray}
		where, the branching ratios $\Gamma_{W\to \mu\nu} = 0.105$, $\Gamma_{Z\to \nu_\mu\bar{\nu}_\mu} = 0.067$. $\alpha_\nu(\alpha_{\bar{\nu}}) = 5.1(9.0)$ while $\beta$ denotes the velocity of the gauge bosons.



\section{Calculations and Results}
In this section, we compute the upper limit of the event rates for neutrino induced muons where neutrinos are originating from the dark matter annihilation in the Sun. For this purpose, we use Eqs. (\ref{eq:Gamma} - \ref{eqnzz}). Two cases are considered here, one is $E_{\rm inj} = m_{\chi}$ and the other is when $E_{\rm inj} = m_{\chi}/3$. For $\nu\bar{\nu}$ production from the dark matter annihilation in the Sun, the channel $\chi\bar{\chi}\rightarrow b\bar{b}$ is considered dominant when $m_b < m_{\chi} < m_W$ ($m_b$ and $m_w$ are masses of $b$ quark and $W$ boson,  respectively). For $\tau\bar{\tau}$ channel as well as for $W^+W^-$ and $ZZ$ channels we consider the regions $m_W < m_{\chi} < m_t$ and $m_{\chi} > m_t$, respectively where $m_t$ denotes the top quark mass.

\begin{figure}
	\centering
	\begin{tabular}{cc}
		\includegraphics[width=0.5\linewidth]{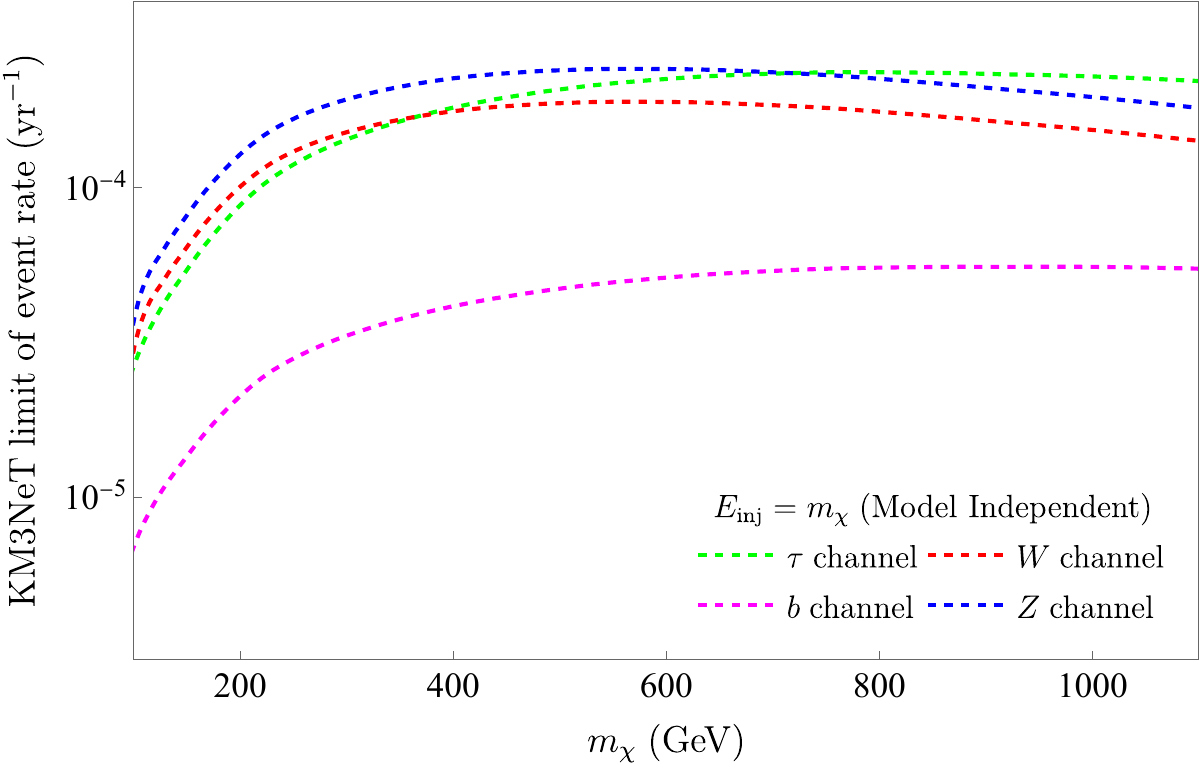}&
		\includegraphics[width=0.5\linewidth]{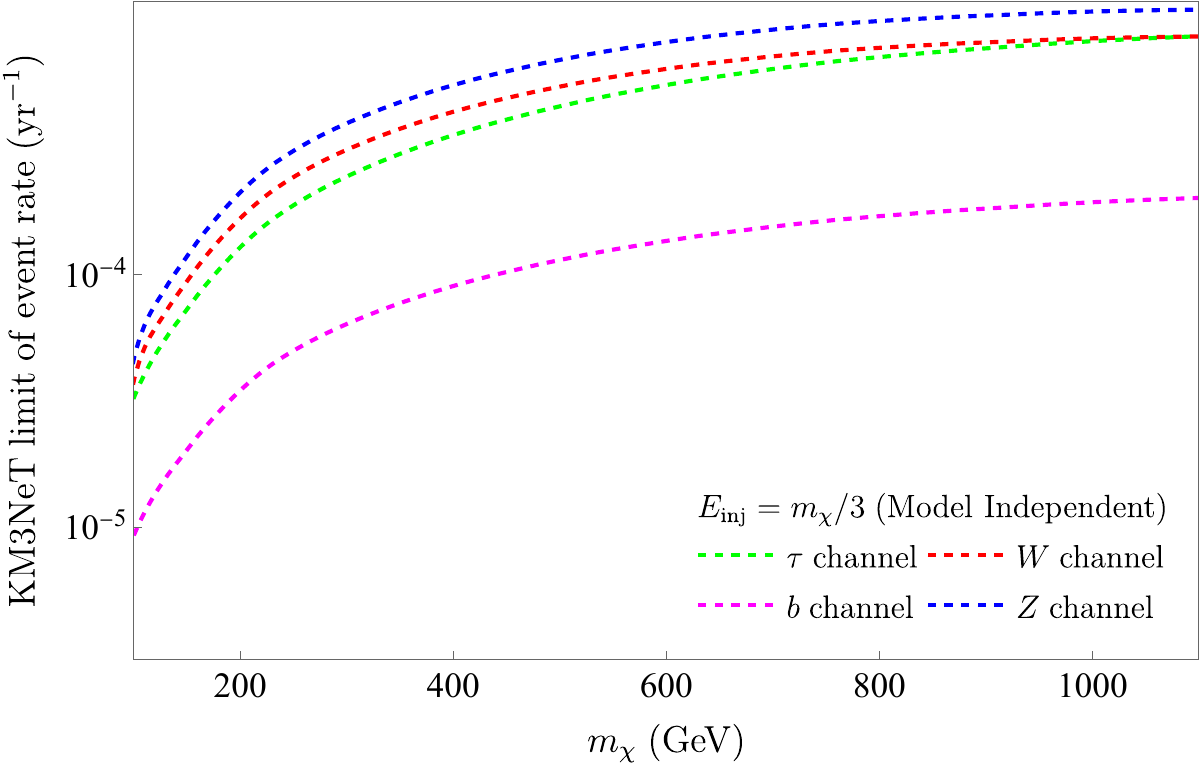}\\
		(a)&(b)\\
		\includegraphics[width=0.5\linewidth]{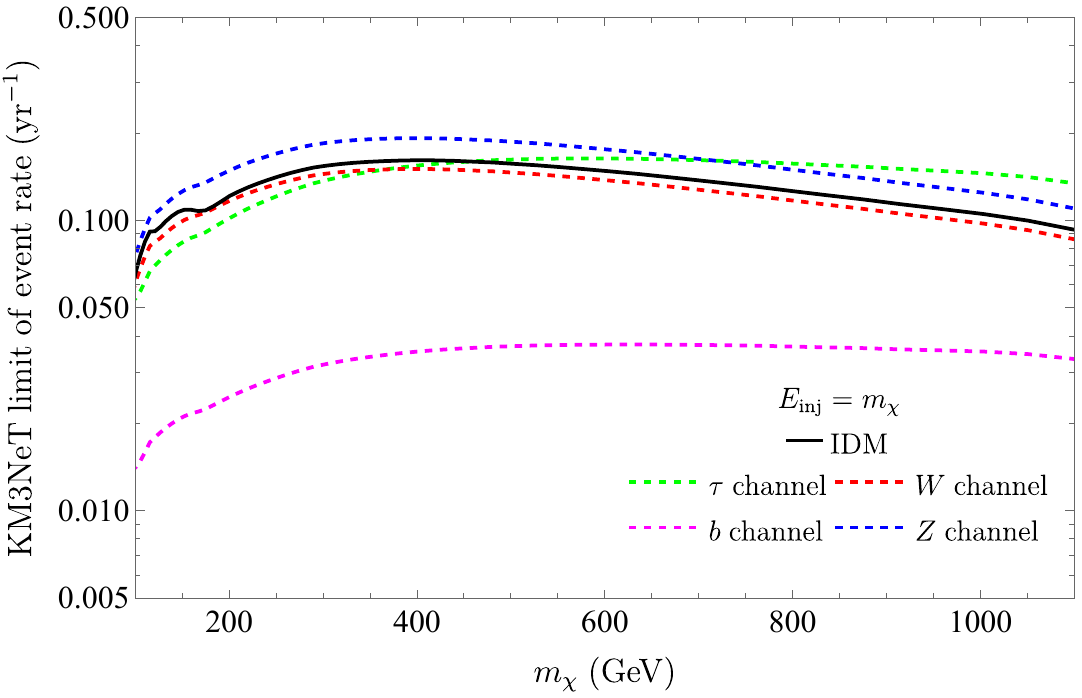}&
		\includegraphics[width=0.5\linewidth]{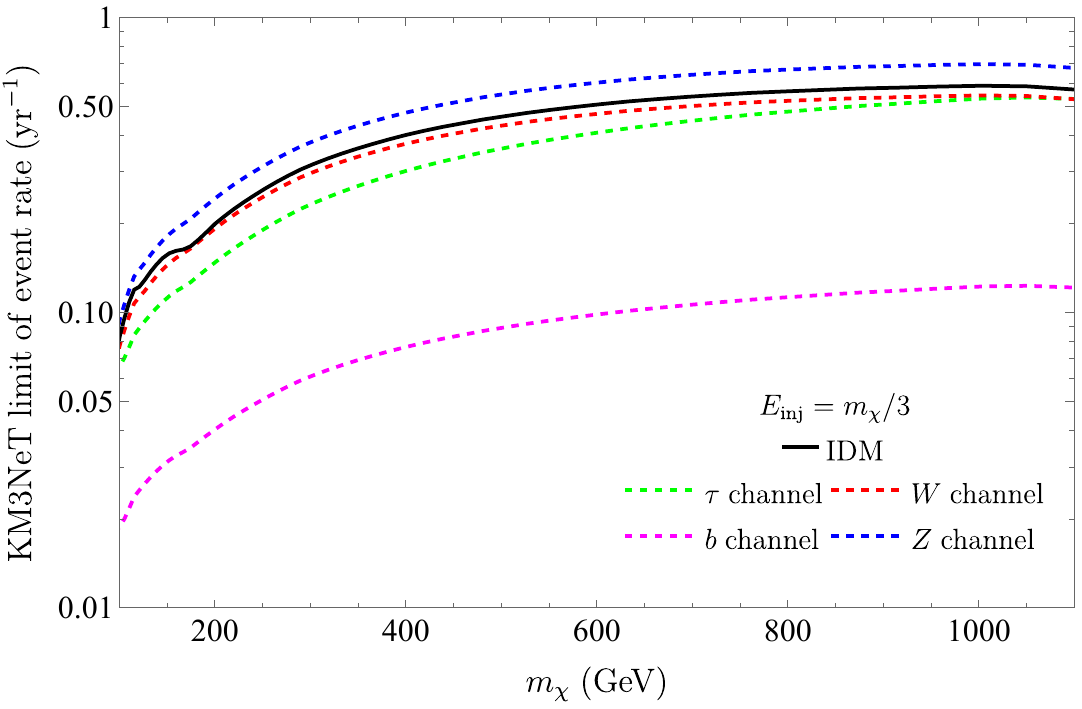}\\
		(c)&(d)\\
	\end{tabular}
	\caption{\label{fig:km3old} KM3NeT upper bounds on WIMP induced muon event rates in different annihilation channels ($\tau$, $b$, $W$, $Z$) with $E_{\rm inj} = m_{\chi}$ (left panel) and   $E_{\rm inj} = \dfrac{m_{\chi}}{3}$ (right panel). In Fig. \ref{fig:km3old}(a) and Fig. \ref{fig:km3old}(b) the results are shown when model independent dark matter is considered while the results when IDM dark matter is adopted are shwon in Fig \ref{fig:km3old}(c) and Fig. \ref{fig:km3old}(d).}
\end{figure}

In Fig. \ref{fig:km3old} we furnish the results for the estimated upper limits for muon event rate per year in the KM3NeT detector for two choices of injection energy $E_{\rm inj}$, namely $E_{\rm inj} = m_{\chi}$ and $E_{\rm inj} =  \dfrac{m_{\chi}}{3}$. We first consider a model independent scenario and obtain the neutrino induced  muon event rates upper limit for four dark matter annihilation channels  $\bar{\chi}\bar{\chi}\rightarrow (\tau\bar{\tau}$, $b\bar{b}$, $WW$, $ZZ)$. These are plotted in Fig. \ref{fig:km3old}(a) and Fig. \ref{fig:km3old}(b) for the cases $E_{\rm inj} = m_\chi$ and $E_{\rm inj} = \dfrac{m_\chi}{3}$, respectively. Note that for the model independent case, the velocity averaged annihilation cross-sections ($\langle \sigma v \rangle$) are computed following the formalism given in the \hyperref[app_sigmav]{Appendix A}.
We have also considered a particle dark matter candidate arising out of a particle dark matter model, namely the Inert Doublet Model (IDM) which is constructed by extending the Standard Model (SM) with an additional SU(2) inert doublet. A $\mathbb{Z}_2$ symmetry imposed on this model under which SM is $\mathbb{Z}_2$ even and the added SU(2) doublet is $\mathbb{Z}_2$ odd. Also the added doublet does not acquire any VEV. These conditions provide the stability to the dark matter candidate in this model (which is the lightest of the two neutral components after the spontaneous symmetry breaking) and it does not generate any added fermion mass via Yukawa interactions. The KM3NeT upper limit for this IDM dark matter for the same annihilation channels are shown in Fig. \ref{fig:km3old}(C) and \ref{fig:km3old}(d) for $E_{\rm inj} = m_\chi$ and $E_{\rm inj} = \dfrac{m_\chi}{3}$, respectively. It is to be mentioned that the annihilation cross-sections ($\langle \sigma v \rangle$) for IDM dark matter (model dependent case) are computed using the \texttt{MicrOMEGAs} code \cite{micromegas_5_2,micromegas_5_0,micromegas_4_3,micromegas_4_1} for IDM dark matter model. The mass of the dark matter is varied from $\mathcal{O}$(100 GeV) to $1.1$ TeV for all the cases. For each of the four annihilation channels, the branching fraction is taken to be 1 in the case of model independent dark matter. However, for the case of IDM dark matter annihilating inside the Sun the branching fraction of each of those channels is separately computed and then the event rates corresponding to each of these channels are added to obtain the upper limit of the total muon event rate at KM3NeT. It is worth mentioning here that we have used the spin independent DM-nucleons cross-section formula as given in Ref. \cite{JUNGMAN1996195} (also see Eq. 16 of \cite{Zentner:2009is}) in all the calculations unless otherwise mentioned.

\begin{figure}
	\centering
	\includegraphics[width=0.7\linewidth]{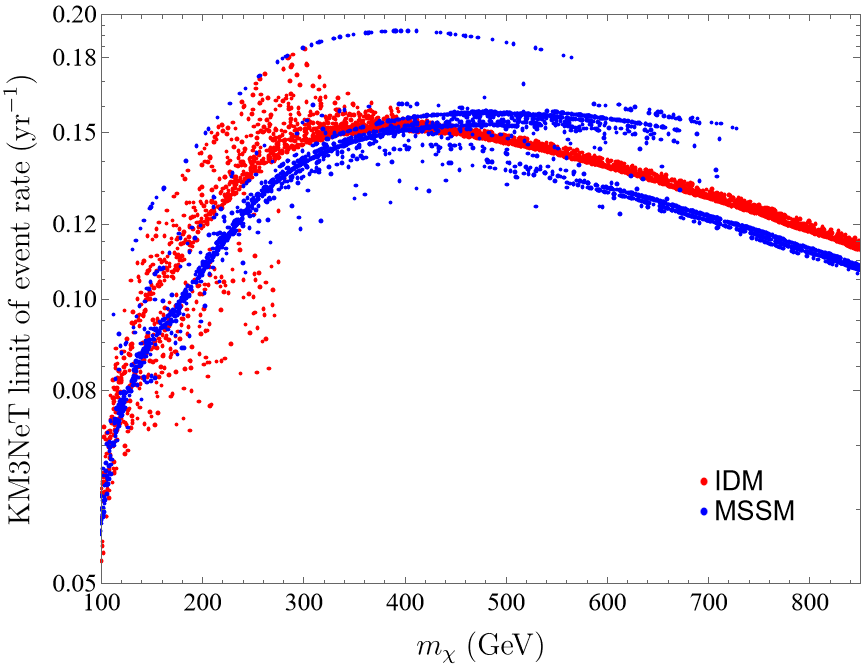}
	\caption{The maximum possible counts for 10000 events are plotted, where dark matter mass and the coupling parameters are randomly chosen in each case. See text for details.}
	\label{fig:km3_idm_mssm_old}
\end{figure}

We then consider the allowed ranges of each of the IDM parameters \cite{LopezHonorez:2010eeh,Barbieri:2006dq} and for each of the chosen dark matter mass, all the parameters are varied within their respective ranges and upper limit of muon event rate for KM3NeT is calculated. The scattered plot thus obtained is shown in Fig. \ref{fig:km3_idm_mssm_old}. We also adopt another popular particle dark matter candidate, namely neutralino which is the Lightest Supersymmetric Particle (LSP) in the supersymmetric theory of Minimal Supersymmetric Standard Model (MSSM) \cite{JUNGMAN1996195} and repeat similar calculations to obtain muon upper limits at KM3NeT. A neutralino in MSSM is the lightest eigenstate of the linear superposition of fermionic super partners of Standard Model gauge bosons and Higgs bosons. A symmetry called R-parity ensures the stability of this dark matter candidate (LSP). As in the case of IDM, here too the MSSM parameters are varied within the allowed ranges \cite{Djouadi:2002ze,Belanger:2008sj,Belanger:2010gh} for each of the chosen neutralino dark matter mass. The annihilation cross-sections for neutralino dark matter for all the four channels mentioned earlier are computed using the \texttt{MicrOMEGAs} code \cite{micromegas_5_2,micromegas_5_0,micromegas_4_3,micromegas_4_1}. The branching fractions of all the four channels considered here (which the annihilating dark matters primarily produce) for each set of parameters with a particular mass are then computed. The KM3NeT limit is then obtained by summing over the upper bounds of the rates for each of the channels considered. These results are also shown in Fig. \ref{fig:km3_idm_mssm_old} with blue scattered plots. It can be noticed from Fig. \ref{fig:km3_idm_mssm_old} that the KM3NeT limit of event rates are in the same ballpark of the results obtained when neutralino dark matter is considered. It can also be seen from Fig. \ref{fig:km3_idm_mssm_old} that the scattered nature of the plot reduces to sharper variations for both the cases beyond $m_{\chi} \approx$ 400 GeV. It appears from Fig. \ref{fig:km3_idm_mssm_old} that KM3NeT limit of event rate is by and large independent of a dark matter model. 

\begin{figure}
	\centering
	\includegraphics[width=0.7\linewidth]{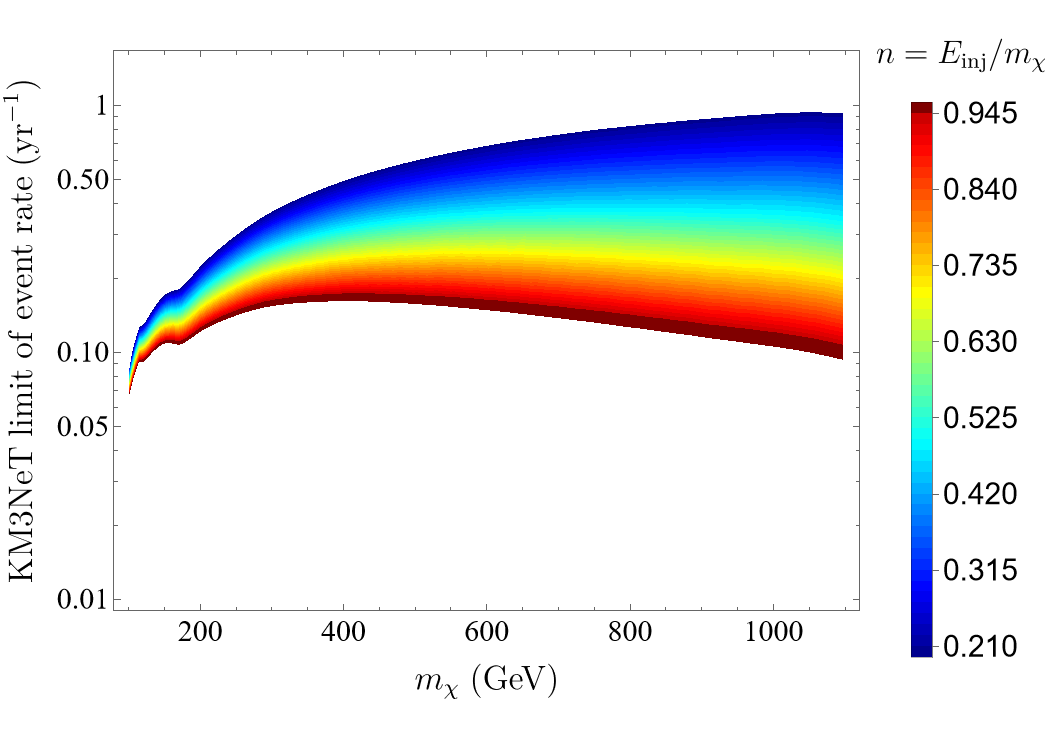}
	\caption{KM3NeT upper bounds on WIMP induced muon event rates for IDM dark matter annihilation, where different forms of $E_{\rm inj}$ are considered. The upper bounds for different $E_{\rm inj}/m_{\chi}$ are represented by the different colour, as described in the colourbar.}
	\label{fig:km3_contour_old}
\end{figure}

In Fig. \ref{fig:km3_contour_old} the KM3NeT upper limit for event rate is plotted for IDM dark matter. Here we parameterize the injection energy $E_{\rm inj}$ by a proportionality constant $n$ as $n=E_{\rm inj}/m_{\chi}$. In this particular case, the branching fraction for different channels are calculated using the \texttt{MicrOMEGAs} code \cite{micromegas_5_2,micromegas_5_0,micromegas_4_3,micromegas_4_1}. The injection energy $E_{\rm inj}$ is continuously varied from 0.1$m_{\chi}$ to 1$m_{\chi}$ in order to obtain the colour plot. The IDM parameters are also varied within the allowed ranges. In this figure (Fig.~\ref{fig:km3_contour_old}), different colours represents the different chosen values of $n=E_{\rm inj}/m_{\chi}$. The value of individual colours are described in the colourbar of Fig.~\ref{fig:km3_contour_old}.\\

\begin{figure}
	\centering
	\begin{tabular}{cc}
		\includegraphics[width=0.5\linewidth]{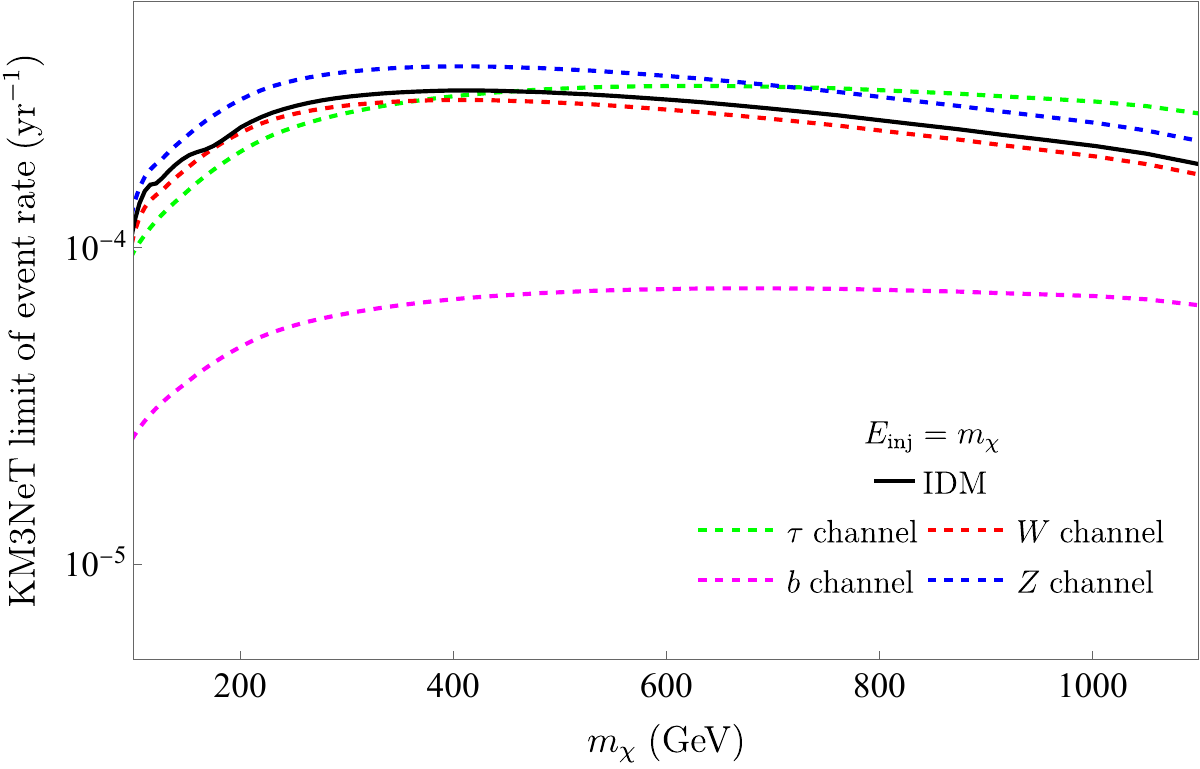}&
		\includegraphics[width=0.5\linewidth]{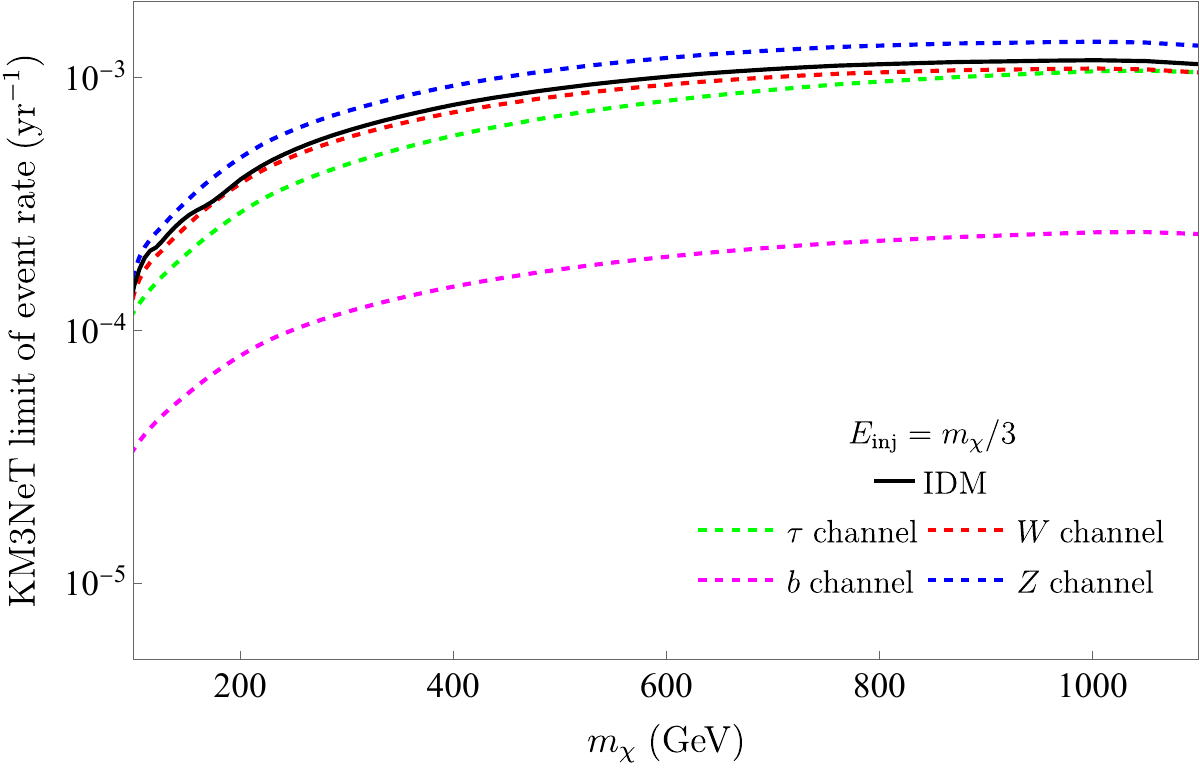}\\
		(a)&(b)\\
	\end{tabular}
	\caption{\label{fig:puu-nu-mat25} The plots are similar to Figs.~\ref{fig:km3old} (c) and (d) but here we use the latest results on WIMPs-nucleons cross-section from PandaX-4T experiment.}
\end{figure}

\begin{figure}
	\includegraphics[width=0.7\linewidth]{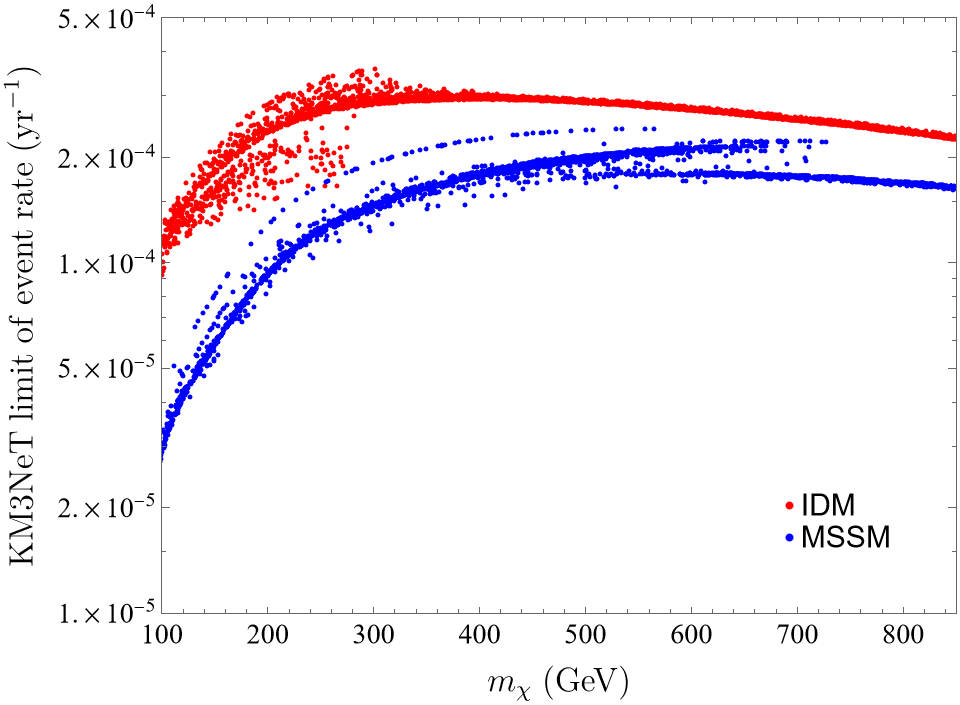}
	\caption{The plot is similar to Fig.~\ref{fig:km3_idm_mssm_old} but using  the latest PandaX-4T results.}
	\label{fig:km3_idm_mssm}
\end{figure}

\begin{figure}
	\centering
	\includegraphics[width=0.78\linewidth]{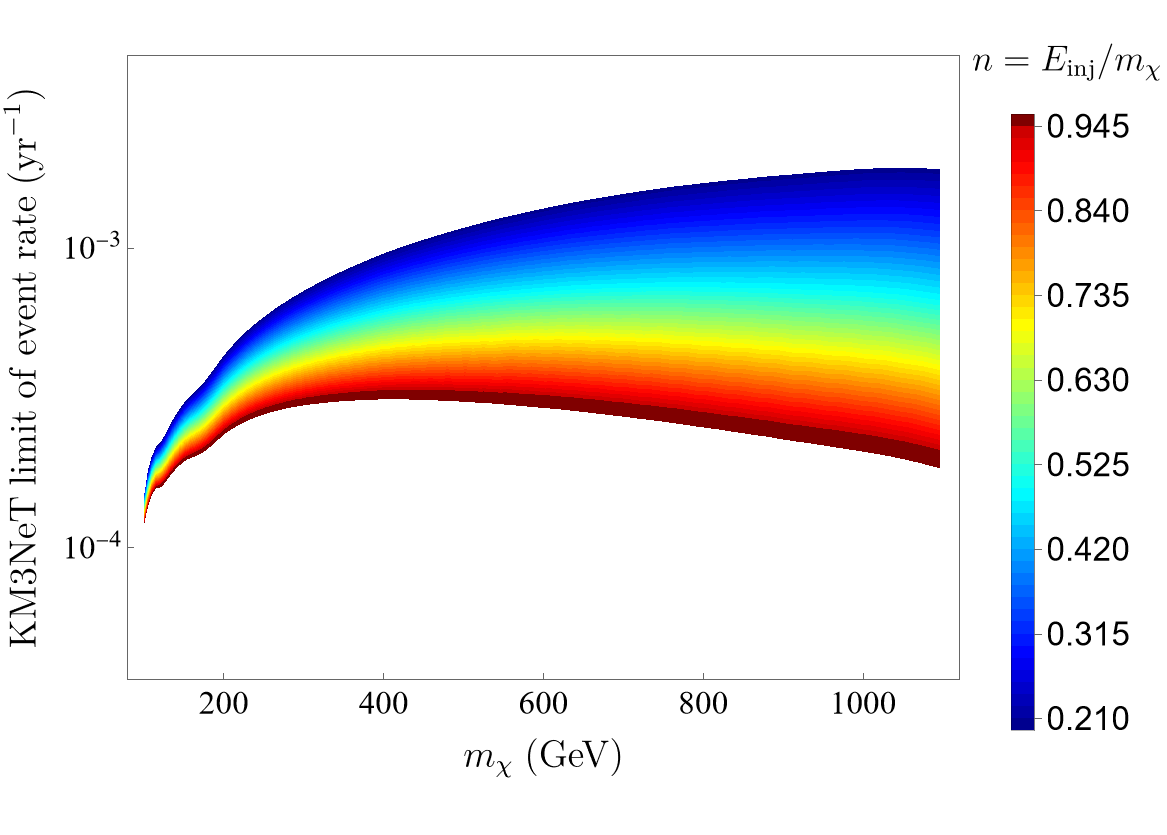}
	\caption{Similar to Fig.~\ref{fig:km3_contour_old} but using the latest PandaX-4T results.}
	\label{fig:km3_contour}
\end{figure}

We then use the latest results on the cross-sections for elastic scattering of WIMPs on nucleons from the PandaX-4T \cite{PandaX-4T:2021bab} experiments and computed the KM3NeT upper limits as is done in Figures \ref{fig:km3old},  \ref{fig:km3_idm_mssm_old}, and \ref{fig:km3_contour_old} to see if any difference in the event rates are observed. We see from Figs. \ref{fig:puu-nu-mat25}, \ref{fig:km3_idm_mssm}, and \ref{fig:km3_contour} that the latest bounds on cross-sections decrease the dark matter annihilated muon event rates by an order of 3. This is largely due to the $\sigma_{\chi}$ limits given by the latest PandaX-4T results.  

\section{ Summary and Discussions}
In this work, we have estimated the upper limit of muon event rates (originating from possible dark matter annihilation in the Sun) for upcoming KM3NeT neutrino telescope.
In an earlier work \cite{Coyle:2009E0}, the author has presented a prediction of muon event rates considering dark matter annihilation, where dark matter is considered in minimal Supergravity (mSugra) and Kaluza-Klein models. But here the calculations are performed using two very famous particle dark matter models, namely IDM and MSSM. For our analysis the annihilation channels that we consider here are, $\tau\bar{\tau}, ~b\bar{b}, ~W^+W^-, \rm {and} ~Z\bar{Z}$. We have found that if we utilize the latest results on the WIMPs-nucleons elastic scattering cross-section from the PandaX-4T experiment, the event rates are decreased by an order of 3. We also observe that the KM3NeT upper limit is more or less independent of a dark matter model. In this analysis, we have found that the event rates are quite small when estimations are made using the DM-nucleons scattering cross-section values as obtained from PandaX-4T experimental results. The detector exposure time in this scenario  would then be very large (almost 1000 years for one event). It however, appears that for particle DM candidates in models such as IDM or MSSM where both the annihilation and scattering cross-sections are calculated within the framework of such models, the muon event rates were found to be much higher. In the latter case, it may take about 10 years for an event to be detected. Enhancement in the detector dimension also raises the effective area and hence the muon event rates at the detectors.          
\section*{Acknowledgements}
A.G. would like to thank Dr. Amit Dutta Banik for important discussion related to the correct form of dark matter capture rate. One of the authors (A.H.) wishes to acknowledge the support
received from St. Xavier’s College, Kolkata and the University Grant Commission (UGC) of the
Government of India, for providing financial support, in the form of UGC-NET-SRF.
\section*{Appendix A: Calculation of $\langle \sigma v \rangle$} \label{app_sigmav}
\begin{figure}
	\centering
	\includegraphics[width=0.78\linewidth]{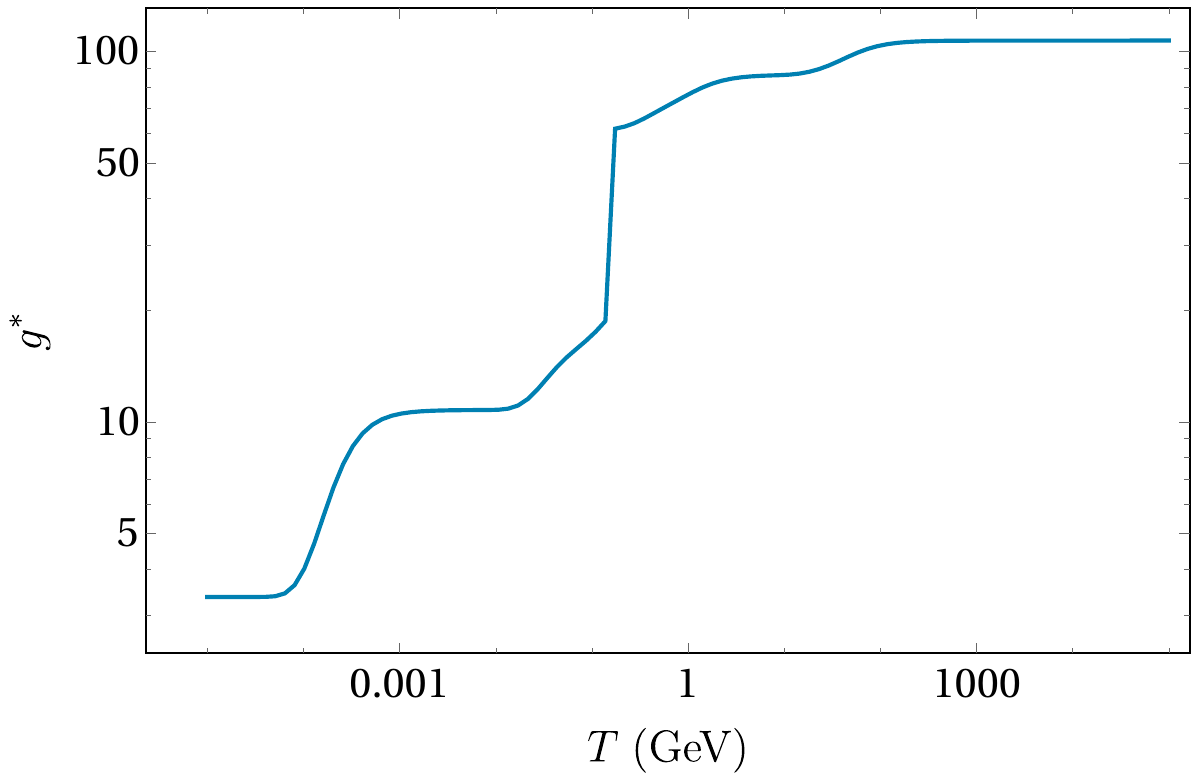}
	\caption{The evolution of relativistic degrees of freedom ($g_*$) for the Standard Model.}
	\label{fig:gstr}
\end{figure}

\begin{figure}
	\centering
	\includegraphics[width=0.78\linewidth]{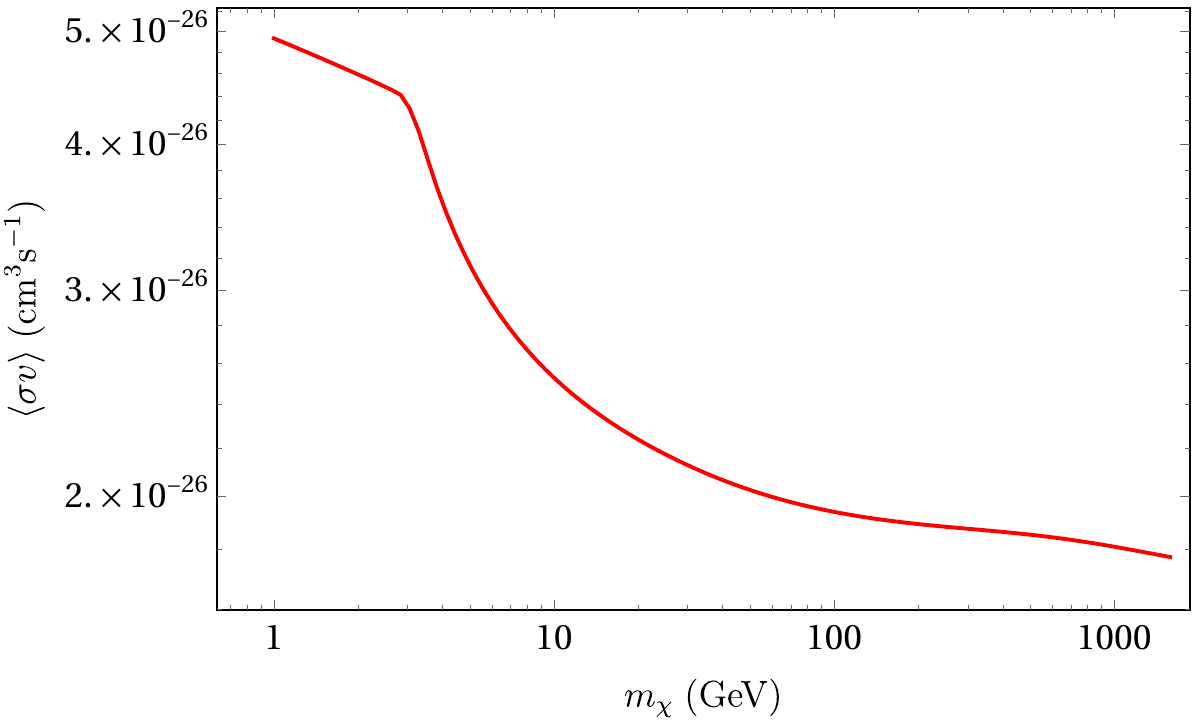}
	\caption{Calculated velocity averaged cross-section of dark matter annihilation.}
	\label{fig:sigmav}
\end{figure}

In order to estimate velocity averaged cross-section for dark matter annihilation (model independently), first one need to evaluate relativistic degrees of freedom ($g_*(T)$) as a function of different temperature (see Fig.~\ref{fig:gstr}), given by
\begin{equation}
	g_*(T) = \sum_{b, f}g_{i}(T),
	\label{eq:gstr}
\end{equation}
where $b$ and $f$ represent all bosons and fermions contribution, respectively  and $g_i(T)$ is the effective degrees of freedom for the i$^{\rm th}$ particle, given by
\begin{equation}
	g_i(T) = \dfrac{15 g_i}{\pi^4}x^4\int_1^{\infty} \dfrac{y(y^2-1)^{1/2}}{\exp(x_i y)+\eta_i}y dy,
	\label{eq:gi}
\end{equation}
where $x_i=m_i/T$, $m_i$ is the mass of the i$^{\rm th}$ particle and $\eta_i$ is a constant, which attains $-1$ for bosons, $1$ for fermions \cite{GONDOLO1991145}.
Now after decoupling, the values of the velocity averaged cross-section (see Fig.~\ref{fig:sigmav}) can be evaluated numerically from the equation (see Eq.~5.5 of Ref.~\cite{GONDOLO1991145})
\begin{equation}
	\dfrac{1}{Y_0}=\dfrac{1}{Y_f}+\left(\dfrac{45}{\pi}G\right)^{1/2} \int_{T_0}^{T_f}g_*^{1/2}\langle\sigma v\rangle dT,
	\label{eq:gondolo_5.5}
\end{equation}
where $T_f$ is the freeze-out temperature.
In the above equation (Eq.~\ref{eq:gondolo_5.5}) the term $\dfrac{1}{Y_f}$ can be neglected \cite{GONDOLO1991145}. The term $Y_0$, present value of the comoving abundance can be obtained from the expression \cite{GONDOLO1991145}
\begin{equation}
	\Omega h^2 \theta^{-3}=2.8282\times 10^8 \dfrac{m}{\rm {GeV}} Y_0,
	\label{eq:y0}
\end{equation}
where $h$ is the Hubble constant in units of 100 km/s/Mpc, $\theta$ is the CMB temperature in units of 2.75 K and $\Omega$ is the present cosmological density parameter which is adopted from the Planck sattelite borne experimental data \cite{planck}.

\section*{Appendix B: IceCube Upper Bounds of Muon Event Rates } \label{sec:icecube}
\begin{figure}
	\centering
	\includegraphics[width=0.78\linewidth]{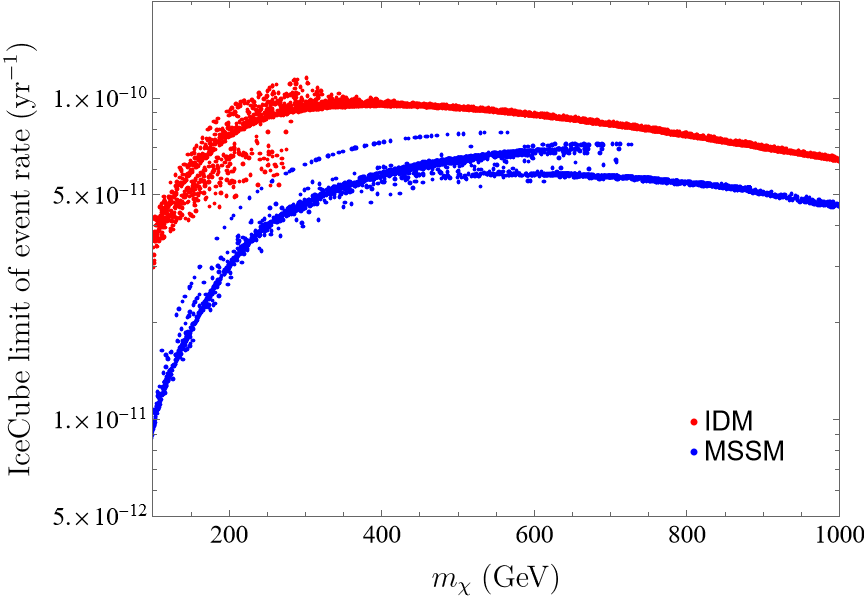}
	\caption{he maximum possible counts for 10000 events are plotted for IceCube neutrino detector, where dark matter mass and
the coupling parameters are randomly chosen in each case.}
	\label{fig:icecube}
\end{figure}

In this section, we have computed the upper bounds of detection rates of Solar neutrinos from annihilation of dark matter at the Solar core considering the IceCube neutrino detector \cite{ic1,ic2,ic3}. The results are shown in Fig. \ref{fig:icecube} for two particular DM models namely IDM and MSSM. In order to compute the event rates we have adopted  the effective area of IceCube detector (A$_{\text{eff}}$) from Fig. 4 of Ref. \cite{Aartsen2017}. We have also found that no significant change is observed in our results even when calculations are made with the IceCube upgraded effective area as given in \cite{Baur:2019jwm}. When comparing with the KM3NeT upper bounds (Fig. \ref{fig:km3_idm_mssm}) it has been  observed that the upper limit of neutrino induced muon event rates are very much suppressed in case of IceCube. This is because of the smaller effective area available for IceCube.

\newpage
\bibliography{DM-references.bib}

\end{document}